\documentstyle[floats,twocolumn,aps,psfig]{revtex}
 
\def\Section#1{}
 
\def\Sz{S^z}

\def\beq{\begin{equation}}
\def\eeq{\end{equation}}
\def\bea{\begin{eqnarray}}
\def\eea{\end{eqnarray}}

\begin{document}
\twocolumn[\hsize\textwidth\columnwidth\hsize\csname 
@twocolumnfalse\endcsname
\title{Ground State Magnetization of Polymerized Spin Chains}
 
\author{Daniel C.\ Cabra$^{1,2}$ and Marcelo D.\ Grynberg$^1$}

\address{$^1$Departamento de F\'{\i}sica, Universidad Nacional de la Plata,
      C.C.   67, (1900) La Plata, Argentina.\\
$^2$Facultad de Ingenier\'{\i}a, Universidad Nacional de Lomas de Zamora,\\ 
Cno. de Cintura y Juan XXIII, (1832), Lomas de Zamora, Argentina}
\maketitle
\date{\today}
\maketitle

\begin{abstract}
We investigate the ground state magnetization plateaus appearing in 
spin $1/2$ polymerized Heisenberg chains under external magnetic fields.
The associated fractional quantization scenario and the exponents 
which characterize the opening of gapful excitations 
are analyzed by means of  abelian bosonization methods.
Our conclusions are fully supported by the extrapolated results
obtained from Lanczos diagonalizations of finite systems.

\vspace{10 pt}
\hspace{1cm} To appear in Phys. Rev. {\bf B}
\vspace{-12 pt}
\end{abstract}
\vskip2pc]


Models of low-dimensional magnets, such as strongly
correlated quantum spin chains \cite{review1} and ladders \cite{review2}, 
are currently receiving  renewed and systematic attention 
for a variety of reasons. Amongst the most remarkable are  
the spin-Peierls dimerization instability \cite{Cross,Castilla},
the Haldane conjecture \cite{Haldane}, and novel concepts such as 
fractional quantization and topological energy gaps \cite{review1,LeH}.
These rather complex phenomena are largely owing to quantum 
fluctuations of individual spins  which tend to restore the rotational 
symmetry of the ground state. Depending on the exchange interactions, 
fluctuations can manifest themselves collectively into
many possible ground states, particularly in 
lower dimensions where their effects are most severe \cite{Sigrist}. 

A wealth of issues have been addressed experimentally to 
confirm these expectations in a series of quasi-one-dimensional
compounds \cite{Chaboussant}. After a vast body of research, 
 it is by now well  established that half-integer-spin chains
are massless whereas integer ones are gapful (see also \cite{CPR}). 
In spite of the availability of a number of excellent realizations 
of  one dimensional Heisenberg antiferromagnets 
\cite{Castilla,Chaboussant}, detailed measurements of spin 
excitations however, have remained confined within the limits of applied 
magnetic fields which are low with respect to the 
exchange interactions thus, precluding a comparison with theory.
Nevertheless, advanced neutron scattering experiments 
spanning large magnetic fields regions 
in relatively low exchange coupling materials, 
e.g. ${\rm Cs_2 Cu Cl_4}$, are now beginning to appear (see e.g.\  
\cite{Coldea}).

Magnetization properties of spin chains have been the subject of 
intensive investigations for quite a while. Important work includes
that of Parkinson and Bonner \cite{PB}, and 
has recently gained new interest 
due to the appearance of magnetization plateaus
studied by several authors in a variety of systems
\cite{Hida}-\cite{Totsuka2}. 
As a contribution to the understanding of massive 
spin excitations in high magnetic fields, here we consider an alternative 
scenario for their low-temperature  realization. 
Specifically, we study the interplay between explicit breaking of
full translational symmetry and applied magnetic fields, 
say along the $z$-direction, 
in spin-$1/2$ Heisenberg chains at $T=0$.
This can be conveniently described  by
a set of Heisenberg antiferromagnets in which the exchange 
coupling interactions  $J_n$ are all equal but one every $p$ sites. i.e.\
\bea
J_n=\left\{
\begin{array}{ll}
J(1-\delta),\ \ \ \ \ \   \ \ \ n/p \in \, {\cal Z}\,,\\
J, \ \ \ \ \ \ \ \ \ \ \ \ \ \ \ \ \ \,\,{\rm otherwise}\,.
\end{array}              \right.
\label{pol}
\eea
It should be stressed that periodic arrays of couplings
are relevant to the study of ferrimagnetic materials \cite{Ferri}
and that also one dimensional dimerized and trimerized 
materials are known to exist \cite{Cross,Hida}.

The Hamiltonians of our polymerized chains in the
presence of a (dimensionless)  magnetic field $h$ applied 
along the $z$-axis are thus given by:
\beq
H_p = \sum_{n=1}^L J_n \,  \vec S_n. \vec S_{n+1}
- \frac{h}{2} \sum_{n=1}^L \Sz_n \,.
\label{1}
\eeq
Here, the  $\vec S_n$ are spin-$1/2$ operators, whereas
periodic boundary conditions are assumed along 
the $L$ sites of the chain  ($\frac{L}{2p} \in {\cal Z}$).
Despite their simplicity, we will show however that these Hamiltonians 
entail a highly non-trivial magnetic behavior controlled solely by the 
chain periodicity $p$ and the external field $h$.

As is known \cite{Griff},  the full translationally 
invariant (FTI) $S=\frac{1}{2}$ chain remains gapless 
for all magnetizations
$\langle M \rangle \equiv \frac{2}{L}\,\langle \sum_n S^z_n \rangle$, 
up  to a saturation field where each of the $L$ individual
spins becomes fully polarized.
On general grounds however, the Lieb-Schultz-Mattis 
theorem \cite{AOY,LSM} indicates that   
FTI Hamiltonians of arbitrary spin-$S$, can be gapful 
provided the magnetization per spin $\langle M \rangle$
satisfies $(S - \frac{1}{2}\langle M \rangle)  \in {\cal Z}$. 
Such gapful excitations should be reflected through the presence of
magnetization plateaus,  in principle at  
these special values of  $\langle M \rangle$.
However, notice that the above theorem {\it does not prove} the
existence of this quantization 
scenario as it refers to non-magnetic 
excitations, i.e.\ modes which preserve the total magnetization.
Nevertheless, magnetization plateaus
have been extensively conjectured and observed 
in both Peierls dimerized \cite{Cross,Totsuka} and trimerized
\cite{Hida} spin chains, 
as well as in frustrated \cite{Tone1,Totsuka2}, anisotropic spin-1
systems \cite{Tone2} and ladders models \cite{CHP}.
They all are examples of a rather subtle phenomenon namely, 
fractional quantization of a macroscopic physical quantity under external 
varying fields. 
Here, we examine this situation for the whole class 
of non-homogeneous chains (\ref{1}), 
attempting to extend and systematize aspects of  quantization
emphasized in those studies.

Following a recent analysis discussed as in \cite{Totsuka,CHP,Totsuka2}
and comprehensively accounted in \cite{LeH,SchBo},
we will apply the by now standard method of abelian bosonization to  
Eq.\ ({\ref{1}).
It is well known that the low-energy properties  of the FTI Heisenberg chain, 
($\delta = 0$), are described by a $c=1$  conformal field theory of a free 
bosonic 
field compactified at radius $R$ for any given magnetization $\langle M \rangle$ 
(see e.g.\ \cite{HaldXXZ}\,). 
The functional dependence of $R$ can be obtained from the
exact Bethe Ansatz solution by solving a set of differential equations
obtained in \cite{bookK} and \cite{Bogo} (for a fuller derivation 
consult for instance Ref. \cite{CHP}\,).
Exploiting this knowledge, the bosonized expression of the low-energy effective
Hamiltonian (\ref{1}) in the homogeneous case $\delta=0$,  can be readily
shown to adopt the form
\beq
\bar{H}= \int {\rm d}x {\pi \over 2} \left\{ \Pi^2(x) +
    R^2(\langle M \rangle) \left(\partial_x \phi(x)\right)^2
\right\}\,,
\label{2}
\eeq
with $\Pi = {1 \over \pi} \partial_x \tilde{\phi}$, and
$\phi = \phi_L + \phi_R$, $\tilde{\phi} = \phi_L - \phi_R$.
Here, the effect of the magnetic field $h$  enters through
the radius of compactification
$R(\langle M \rangle)$. This radius governs the conformal
dimensions, in particular the conformal dimension
of a vertex operator ${\rm e}^{i \beta \phi}$ is given by
$\left({\beta \over 4 \pi R}\right)^2$. 
Within the framework of the theory of Luttinger liquids,
it is worth pointing out also that the compactification radius 
is related to the  parameter  ${\rm K}$ by $R^2=\frac{\rm K}{4\pi}$.
 
The bosonized expressions for  the spin operators read:
\beq
S_i^z(x)\approx \frac{1}{\sqrt{2\pi}} \frac{\partial \phi_i}{\partial x}
 +const.  :\cos(2k_F^i x+\sqrt{4\pi}\phi_i):+\frac{\langle M\rangle}{2} \, ,
\label{3}
\eeq
and
\beq
S_i^-(x)\approx e^{-i\sqrt{\pi} \tilde{\phi_i}}(1+const.
:\cos(2k_F^i x+\sqrt{4\pi}\phi_i):),
\label{4}
\eeq
where the colons denote normal ordering with respect 
to the ground state with magnetization $\langle M \rangle$.
Now we apply this methodology to compute the effective form of the
interaction. After some algebra, in the limit of weak polymerization
$\delta\ll 1$, it can be readily shown that the most relevant perturbation 
term  is given by
\beq
H_{int}\approx  \delta \sum_{x'=1}^{L/p} \cos(2k_F (px'+\frac{1}{2})
+\sqrt{4 \pi}\,\phi)\,. \label{5}
\eeq
This operator will survive in passing from the lattice to the continuum
model, assuming that the fields vary slowly, only when the oscillating
factor $\exp(i2px' k_{F})$ equals one.  
Since the Fermi level is given by 
$k_F = {\pi \over 2} (1\! - \!\langle M \rangle )$, 
this in turn will happen when the condition
\beq
{p \over 2} (1 - \langle M \rangle) \in {\cal Z} \,
\label{condM}
\eeq
is satisfied . 

We can now study when a plateau will appear in the magnetization curve
for the polymerized chain (\ref{1}).
To do that, we first have to see which are the values for the magnetization
where there could be a plateau,
for a given value of the period $p$, (i.e.\ solve for (\ref{condM}))
and then we need to evaluate the scaling dimension
of the operator (\ref{5}), which at zero loop  is given by
\beq
d=\frac{1}{4\pi R^2},
\label{dim}
\eeq
which is in turn governed by the radius of compactification as we already stressed. 
By virtue of the lower bound of the compactification radius  \cite{bookK,Bogo}, namely
$R(\langle M \rangle) \ge R(\pm 1)=1/(2\sqrt{\pi})$ , it follows from Eq.\,(\ref{dim})
that $d < 2$ for all magnetizations $\vert \langle M \rangle \vert < 1$. 
This ensures the {\it relevant} character of the operator (\ref{5}) which in turn survives
in the continuum limit  whenever Eq.\,(\ref{condM}) is satisfied. Therefore, 
we can conclude that  constraint 
(\ref{condM}) constitutes a {\it sufficient condition} ultimately responsible for
the appearance of magnetization plateaus and massive spin excitations. 
This is our main result.

We now turn to a numerical finite-size analysis. In  Figs.\,1(a)--(e) we display 
a variety of magnetization regimes as a function of both polymerization 
parameters $J'/J \equiv 1-\delta$ and applied magnetic fields $h$.
This is a rather compact form of representing conventional magnetization curves 
for different polymerization strengths. Here, each line is associated 
to successive values of  $\langle M \rangle$ which decrease monotonically 
from top to bottom, as they should for a non-frustrated system.
The results were obtained from exact diagonalization of finite systems 
via a recursion type Lanczos algorithm \cite{Lanczos} applied on each
magnetization  subspace $S^z = \{ 0,1,\,...\,,L/2\}$.  To avoid
the formation of spurious interfaces, even multiple lengths
of the lattice periodicity were taken throughout.
Using fully isotropic chains up to $L = 24$ sites  with periodic 
boundary conditions,  our numerical analysis supports
entirely the quantization constraint (\ref{condM}). 

As expected,  the ground state 'phase 
diagrams' exhibit  bands of empty regions corresponding
to the magnetization plateaus of $S^z$ 
referred to above, while regions filled
with magnetization lines reflect smooth magnon 
excitations arising in the thermodynamic limit $L \to \infty$.
It can be readily observed that
for chains of periodicity $p > 1$ (dimers, trimers, etc),  
a plateau-like structure emerges precisely at the {\it rational} 
magnon densities $\langle M \rangle = 1 - 2 q /p$, 
$(q=0,\, 1, \,...\,p\,)$, implicit in the general scenario of 
Eq.\,(\ref{condM}). It is worth remarking on the robustness of this 
topological constraint as similar results 
continue to hold for anisotropic ($X\!X\!Z$) chains, the plateaus
always appearing at the {\it same} values of $\langle M \rangle$.

We can also predict the behavior of the mass gap (width of the plateau),
with the polymerization strength  $\delta$ by means of a simple zero loop 
computation \cite{ASZX}. 
Aside logarithmic corrections to the case
$\langle M \rangle = 0$, this yields
\beq
{\rm g} \propto \delta^{1/(2-d)}\ ,
\label{gap}
\eeq
with $d$ given as in (\ref{dim}).

To enable an independent check of this result,  
we now turn to  the issue of extrapolating 
the numerical finite size estimates of the mass gaps ${\rm g}_L$ 
towards their corresponding thermodynamic limits.
Note, on one hand, that {\it any} 
extrapolation procedure by necessity 
assumes that the asymptotic behavior applies 
to the values of $L$ within reach.
However, it is known \cite{Griff} that finite 
size corrections to the gap in the
excitation spectrum of the homogeneous
Heisenberg chain vary slowly 
as  ln (ln\,$L$)/ln$^2 (L)$, thus affecting 
the results over a wide range of sizes.
In fact,  as can be seen in Figs. (\ref{figure0}),  
this turns out  to be the case also for weak 
polymerization regimes, $\delta \to 0$, where 
finite size effects are more pronounced. Therefore, in studying  
numerically the mass gap behavior obtained in Eq.\,(\ref{gap})
we are confronted to restrict considerations to the non-critical region 
$ 0 < \vert \delta \vert  \le 1$, yet suitable to test independently the 
correctness of our bosonization approach. 

To estimate the actual masses in the limit $L \to \infty$, 
we fitted  the whole set of finite-size results 
(even integer multiples of $p$ within the range $4 \le L \le 24$), 
using  both linear, and logarithmic type methodologies of 
convergence \cite{Gutt}, i.e.\
\begin{equation}
\label{extra}
{\rm g}_L  \simeq {\rm g} \, + \,A  \, e^{-B\,L}\; , \;\;\; 
{\rm g}_L  \simeq {\rm g} \, + \,A /L^{B}\,.
\end{equation}
Either extrapolation procedure yields basically the same result
with at least  3 significant digits. This latter variation ultimately gives an estimative 
idea of the lower bound of the extrapolation error.
The reliability of our results was checked also by comparing the trend
arising from {\it smaller} systems
($L \le 20$). When  the critical region is approached however,  the accuracy 
differs widely, particularly for  $\vert \delta \vert  < 0.2$.

Although there are alternative extrapolation algorithms
which do not involve fits to specific forms  \cite{Gutt} 
we should hasten to add  however, that  their efficiency depends strongly 
on the abundance of data. In our case, this is 
translated in the availability of matching sizes, already constrained by both 
the periodicity $p$ 
and the antiferromagnetism.
Nevertheless, we  were able to find a remarkable agreement with the
compactification radius comprehended in Eq.\,(\ref{dim}) and the 
exponents of  Eq. (\ref{gap}).
The results are shown in Fig.\,2  where 
we display respectively the gap openings
around $\langle M \rangle = $0, 1/3, 1/2,  
for $p=2,\, 3\,$, and 4. The dimerized case reproduces the 
well known $2/3$ exponent predicted in \cite{Cross} and 
corroborated subsequently by diverse numerical studies \cite{Spron}.
To our knowledge however, opening exponents for $\!p \ge 3\!$ 
(see Fig. 2),  have not been elucidated yet by other investigations.

Finally, it is instructive to comment further on the role of
{\it quantum} fluctuations namely, the tendency 
of spins to spontaneously tilt occasionally due to the Heisenberg
uncertainty relations, and their relevance to our results \cite{Sigrist}.
For classical spins, e.g. Ising and $n$-vector
models, the interplay between  dimensionality and {\it statistical} 
fluctuations, though crucial in determining phase transitions,
is not sufficient to entail the fractional behavior studied so far.
In fact, an elementary transfer matrix calculation shows that 
the Ising equivalent of (\ref{1}) wipes out all but
two magnetization plateaus,  namely $\langle M \rangle = 0,\, 2/p$, 
(even $p  > 2$), their widths behaving linearly with $\delta$.
Thus, it is worth pointing out  that Eqs.\ (\ref{condM}) and (\ref{gap}),
in contrast,  constitute a genuine macroscopic quantum effect.

In summary, we have presented a bosonization picture that accounts for 
the fractional quantization observed in a class of non-homogeneous 
Heisenberg antiferromagnets. All low-energy exponents which characterize 
the opening of gapful excitations have been obtained and treated on an 
equal footing while checked with Lanczos diagonalizations of finite systems.
Aside these theoretical pursuits, we trust our study will help to convey a 
clearer understanding of the many characteristics present in {\it real} low 
dimensional magnets. A similar analysis in polymerized ladder systems 
is in progress.

It is a pleasure to acknowledge fruitful discussions with A.\ Honecker,
P.\ Pujol and F. A.\  Schaposnik. The authors acknowledge financial
support of CONICET and Fundaci\'on Antorchas.

\begin{figure}[htp]
\psfig{figure=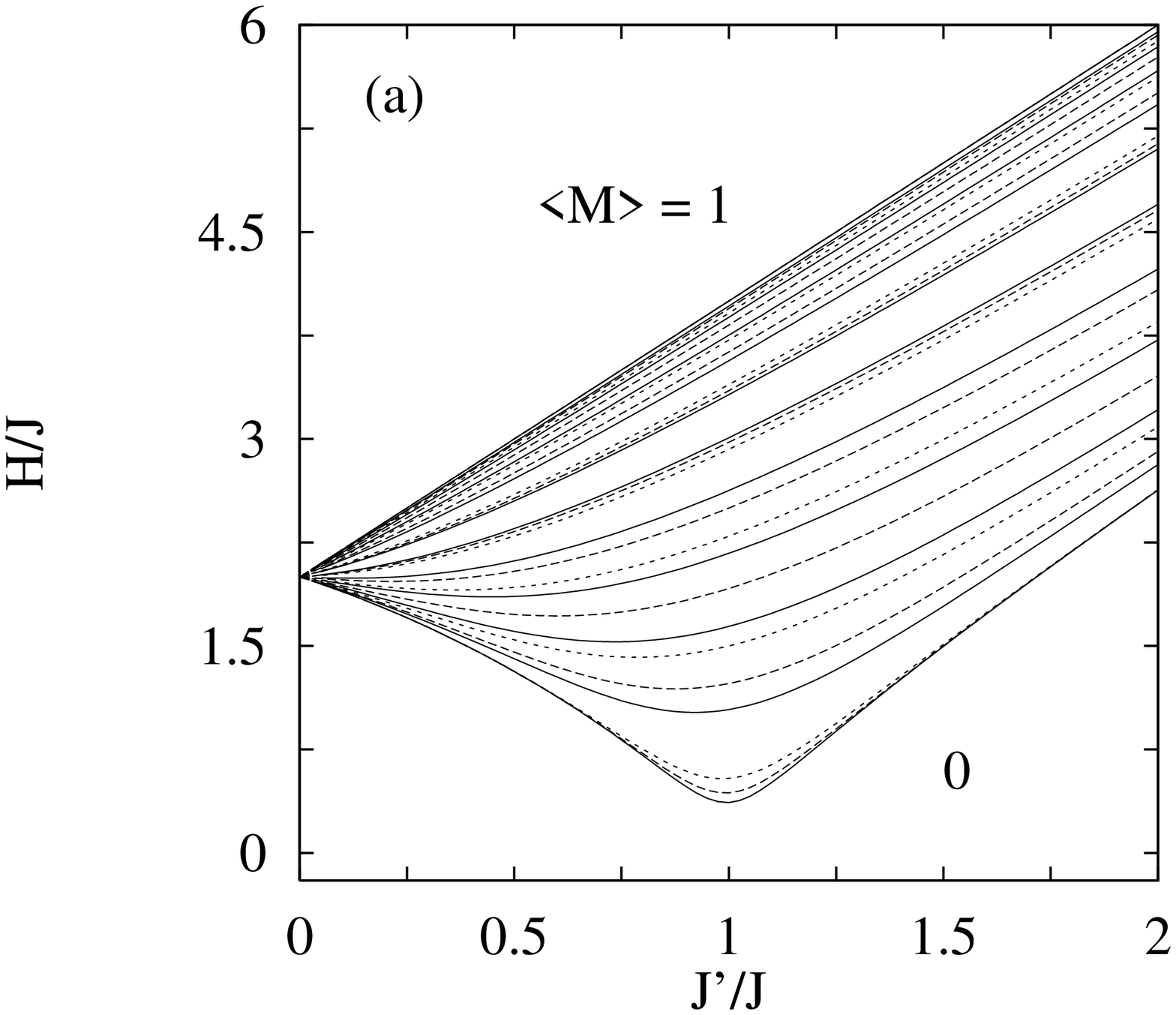,width=1.0\columnwidth,angle=0}\hfil\\
\psfig{figure=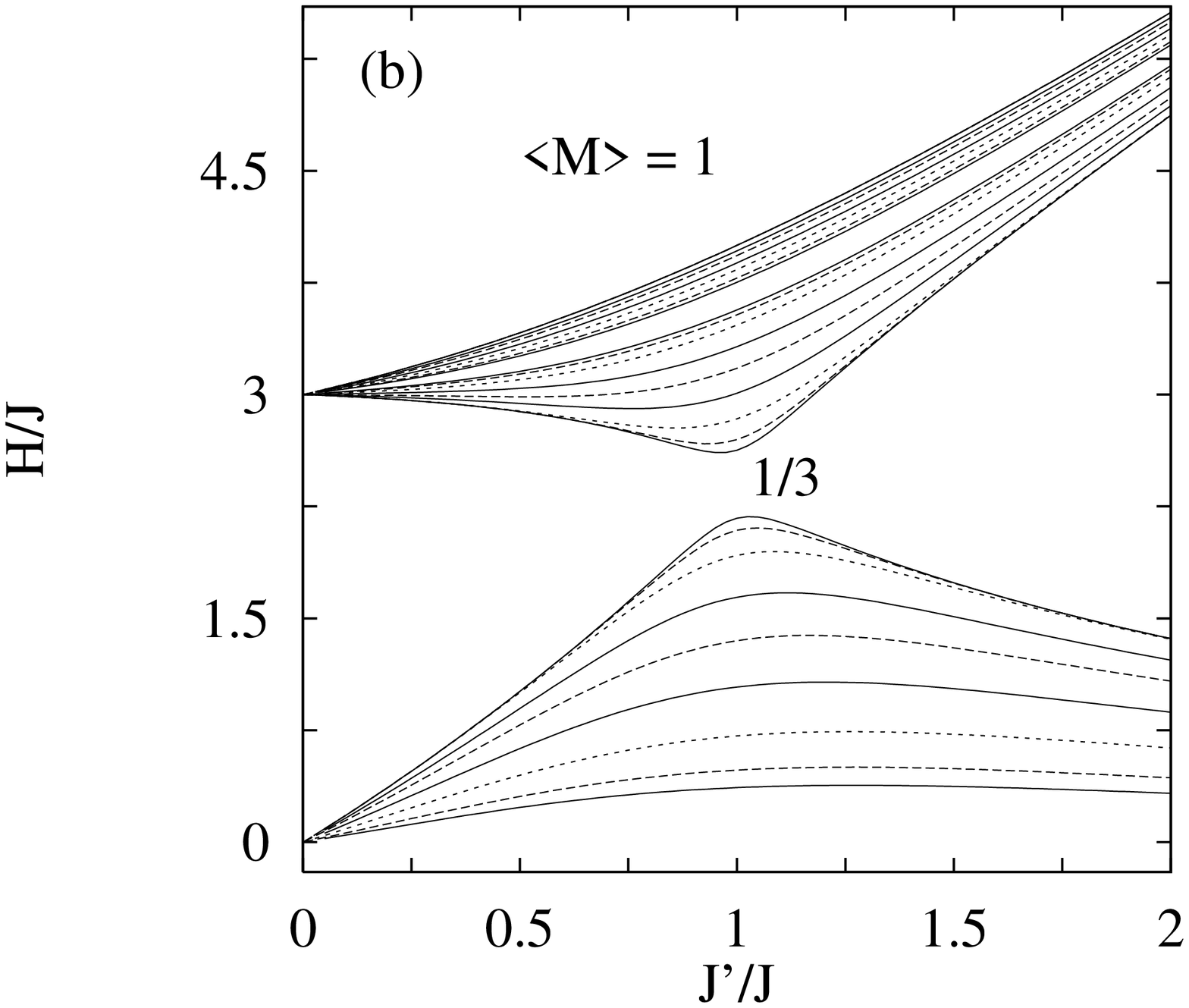,width=1.0\columnwidth,angle=0}\\
\psfig{figure=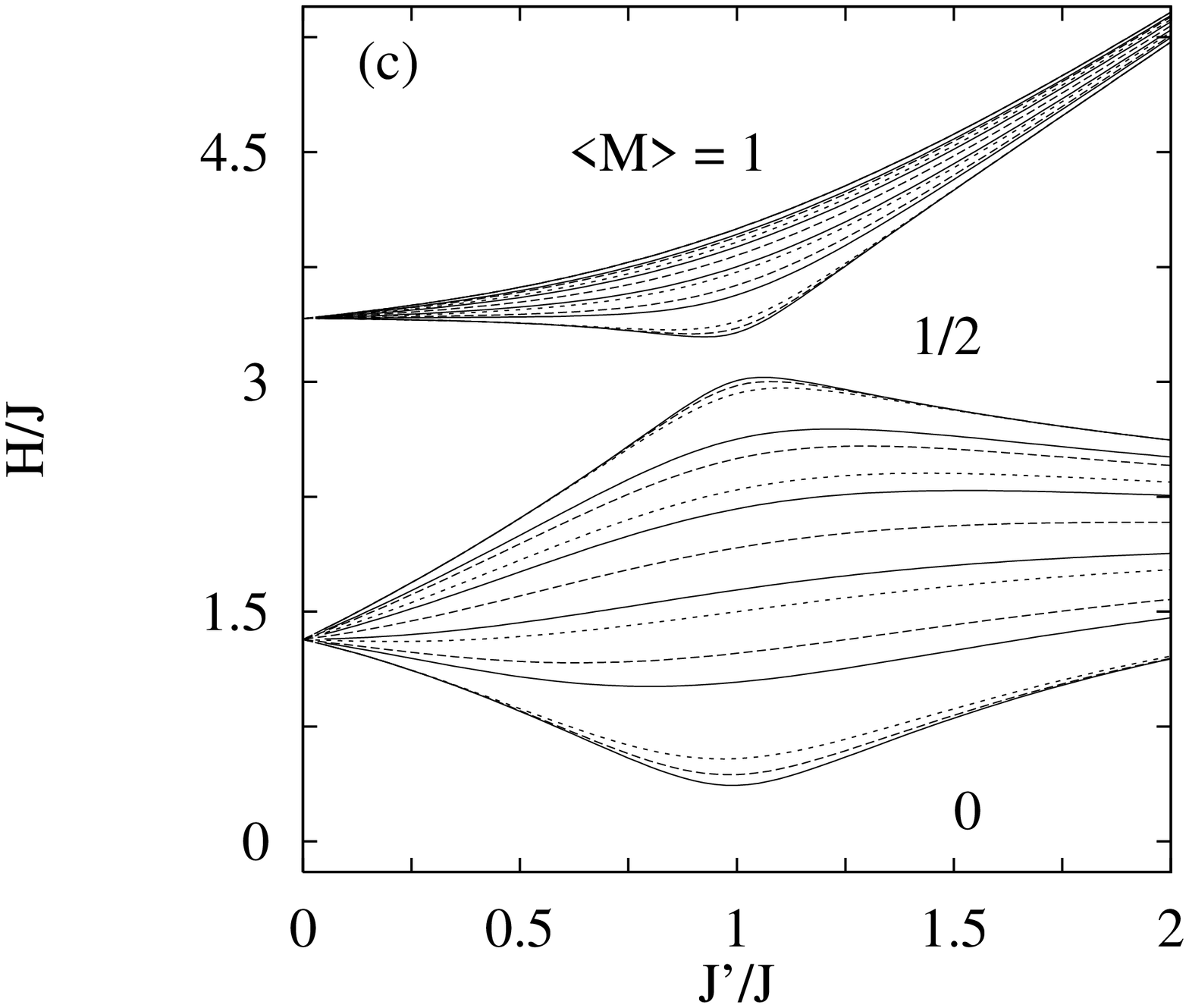,width=1.0\columnwidth,angle=0}
\end{figure}
\begin{figure}[htp]
\psfig{figure=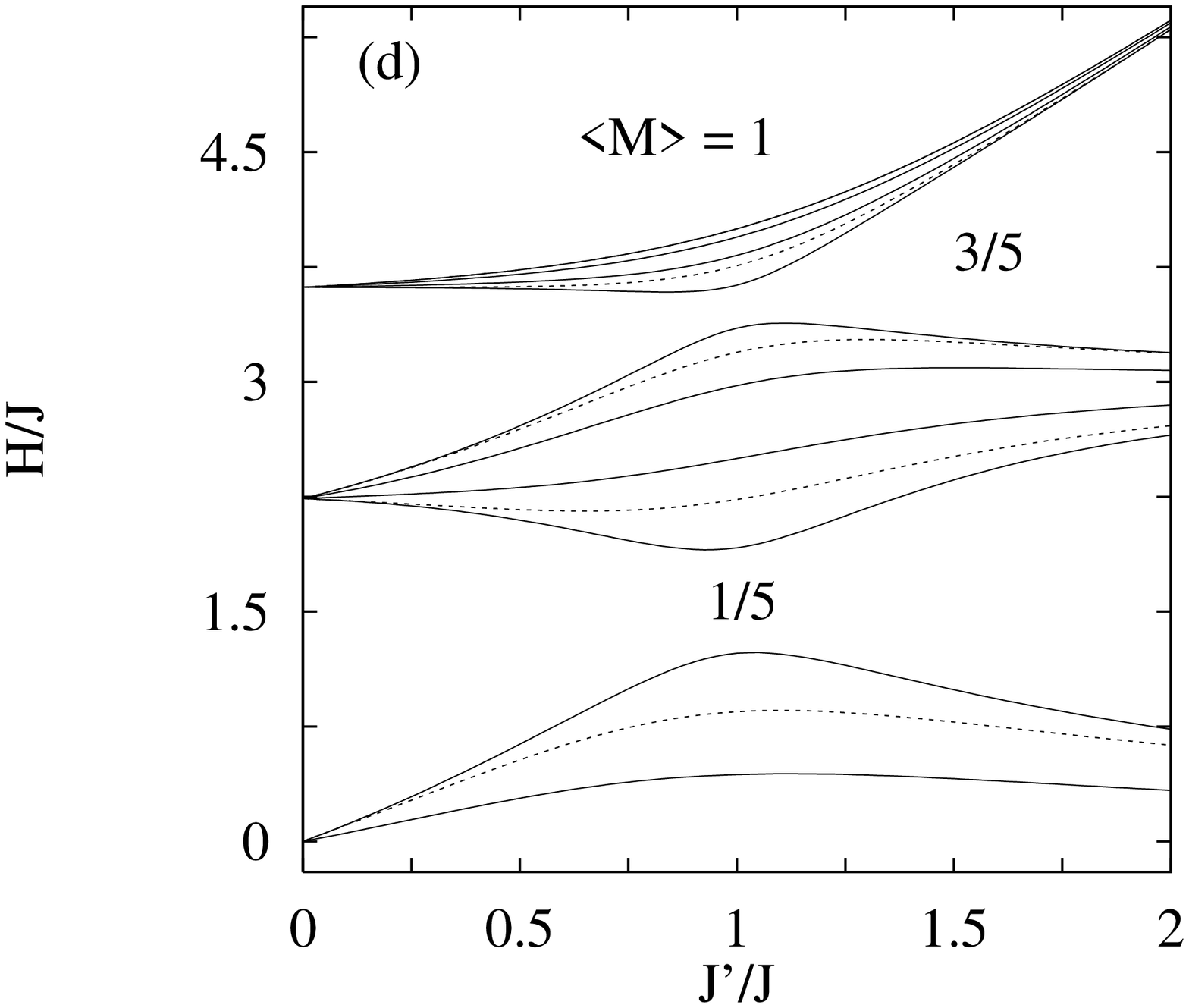,width=1.0\columnwidth,angle=0}\hfil\\
\psfig{figure=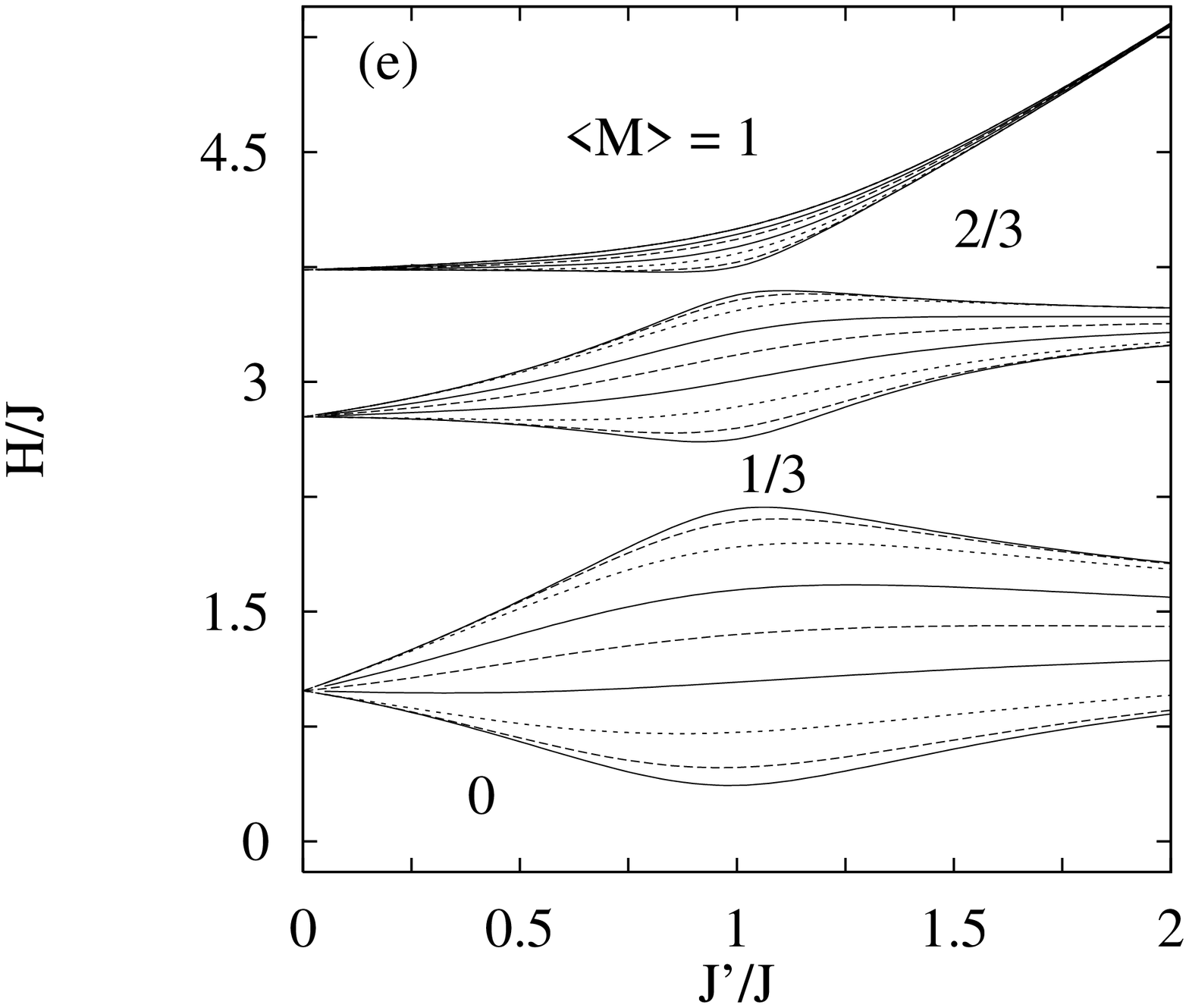,width=1.0\columnwidth,angle=0}
\caption{Magnetization contours of finite 
polymerized chains for  (a) $p=2,\, L=24, 20, 16$\,; 
(b) $p=3,\, L=24, 18,12$\,;
(c)  $p=4,\, L=24, 20, 16$\,; 
(d)  $p=5,\, L=20, 10$ (full and dotted lines respectively)
and; (e) $p=6,\, L=24, 18, 12$\,. Except for (d), 
full, dashed, and dotted lines
stand respectively for large, medium and small sizes.
They denote all accessible magnetizations, whereas their values
decrease from top to bottom. Though numerical 
accuracy in $h/J$ is bounded 
by $10^{-7}$,  size effects become evident for $J'=J$, as no 
plateaus (empty wide bands), 
should occur in the thermodynamic limit.
\label{figure1}
}
\end{figure}

\begin{figure}[ht]
\psfig{figure=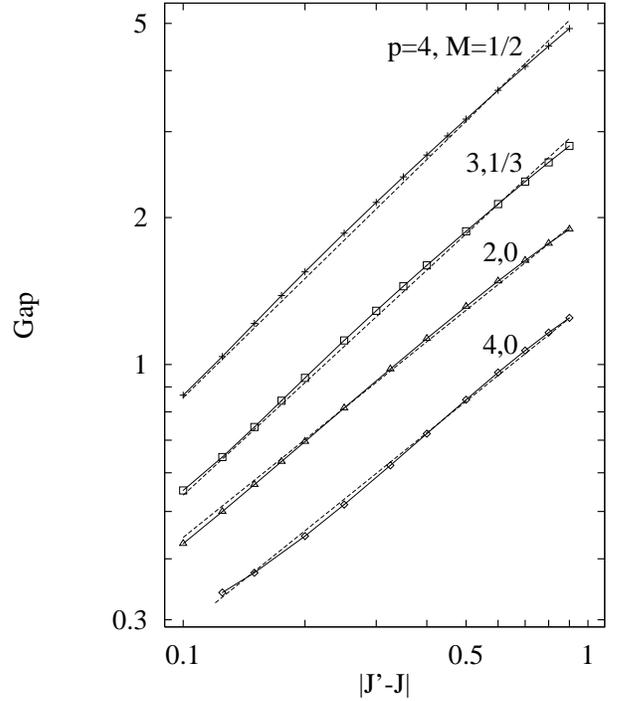,width=\columnwidth,angle=0}
\smallskip
\caption{Extrapolated values of the gap for $p=2$ around
$\langle M \rangle = 0$, $p=3$ for
$\langle M \rangle = 1/3$ and, $p=4$  with 
$\langle M \rangle = 0, 1/2$. 
Solid lines are guide to the eye whereas slopes of dashed lines 
denote the estimated opening exponents,  namely
(in descending order), 0.8(1),  0.77(10) and,  0.66(10),  
($\langle M \rangle = 0$). 
To improve the clarity of the figure, the uppermost curve 
was shifted multiplying the gap by a scale factor 2.5.
\label{figure0}
}
\end{figure}

\end{document}